\documentstyle[11pt,newpasp,twoside,epsf]{article}
\def\edcomment#1{\iffalse\marginpar{\raggedright\sl#1\/}\else\relax\fi}
\reversemarginpar
\begin{document}
\title{Linear Polarization Properties of Pulsars at 35 \& 327~MHz}
\author{Ashish Asgekar$^{1,2}$ \& A. A. Deshpande$^{1}$}
\affil{$^{1}$Raman Research Institute, Sadashivnagar, Bangalore 560 080 INDIA.}
\affil{$^{2}$Joint Astronomy Programme, Indian Institute of Science, Bangalore~560 012 INDIA.}
\vspace{0.5cm}
\indent Faraday Rotation of the plane of polarization of broad-band
signals, during propagation through the intervening medium, manifests as
quasi-sinusoidal spectral modulations when observed with a
telescope sensitive to a single linear polarization. Such
a modulation can be exploited to study linear
polarization characteristics of pulsars (Suleimanova, Volodin,
\& Shitov, 1988; Smirnova \& Boriakoff 1997, Ramkumar \& Deshpande 1999 (RD99)).
We have used our data on a few bright pulsars at 35~MHz
(Asgekar \& Deshpande 1999, elsewhere in this volume)
and data obtained at 327~MHz
using Ooty Radio Telescope (see RD99 for details), to
study average linear polarization properties using this technique.
The data obtained over 256 frequency channels were re-sampled in the
spectral domain to make the Faraday modulation appear periodic,
and then a simple Fourier analysis was performed to look for
(ACF) features associated with the possible spectral modulation
(see RD99 for the analysis details).\\
\indent We show, in fig.~1, the spectral modulation in the case of
\mbox{J0837-4715} at 327~MHz, along with the average polarization behavior.
We estimate 1) the apparent $RM~\simeq{131}~rad/m^{2}$; 2) the percentage linear
polarization~$\simeq{25\%}$. For the \mbox{pulsar J0846-3533}, we find
1) $RM\simeq~126~rad/m^{2}$ and 2) fractional linear
polarization ~$\sim 33\%$ (fig.~2),
whereas the corresponding estimates for \mbox{J0924-5302} are
$\sim 126~rad/m^2$ \& $\sim$15\% respectively. Our RM estimates are consistent
with the catalogued values within the uncertainties in the ionospheric contributions.\\
\indent In the data on B0834+06 at 35 MHz,
we detect a weak but noticeable spectral modulation that corresponds to
a period of~16.1 channel-widths. This would imply an RM~$\simeq{13}$ rad/m$^2$
(the catalogue $RM=23.6~rad/m^{2}$). The average intensity corresponding to this
modulation is shown in fig.~2 (without any statistical correction).
We estimate the percentage polarization to be less than~$\sim{38\%}$.
No periodic spectral modulation feature was obvious
in the 35~MHz data on the pulsars B1133+16 \& B0943+10,
though the spectra did show some quasi-periodic intensity modulations.
\begin{figure}
\plottwo{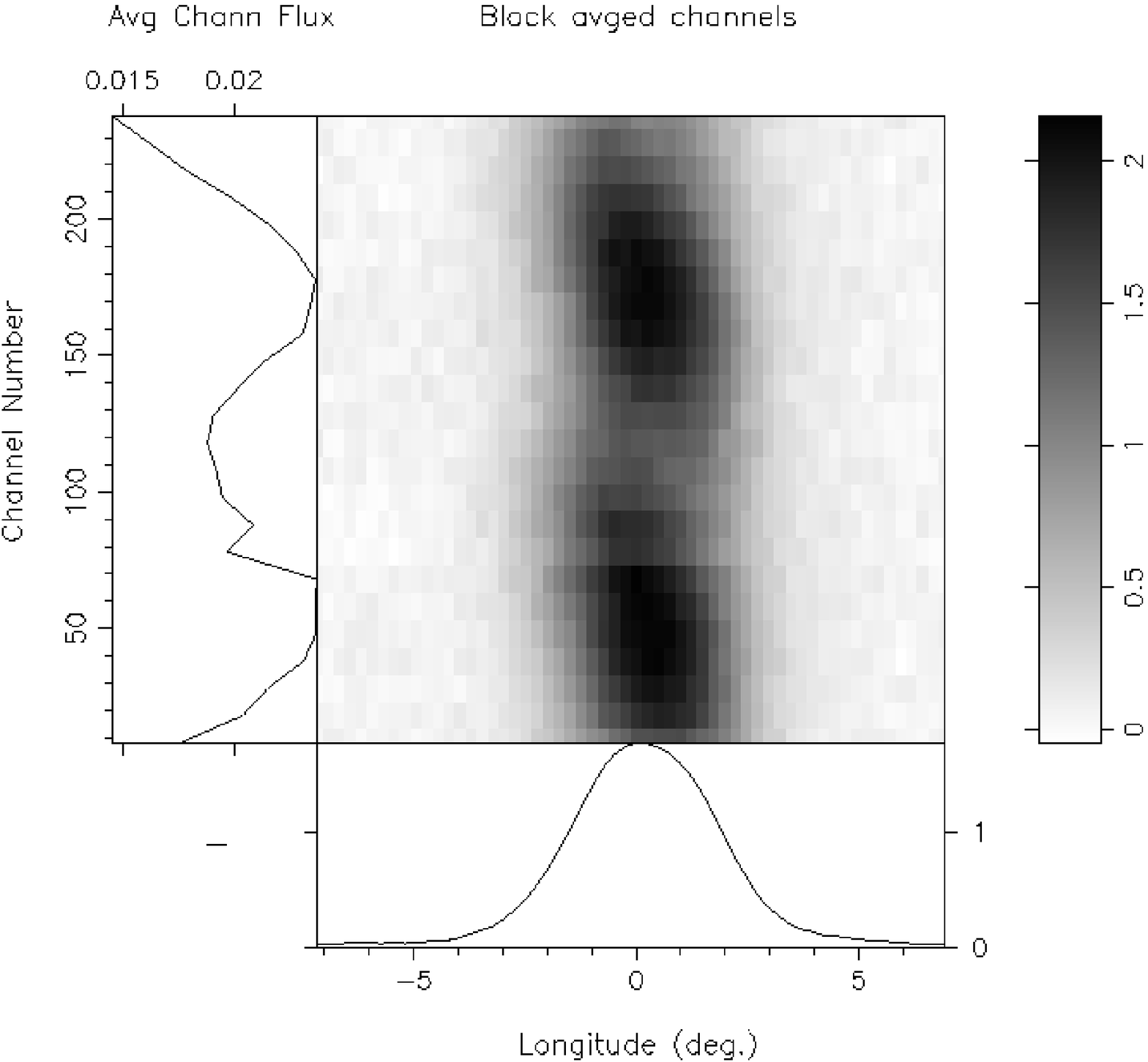}{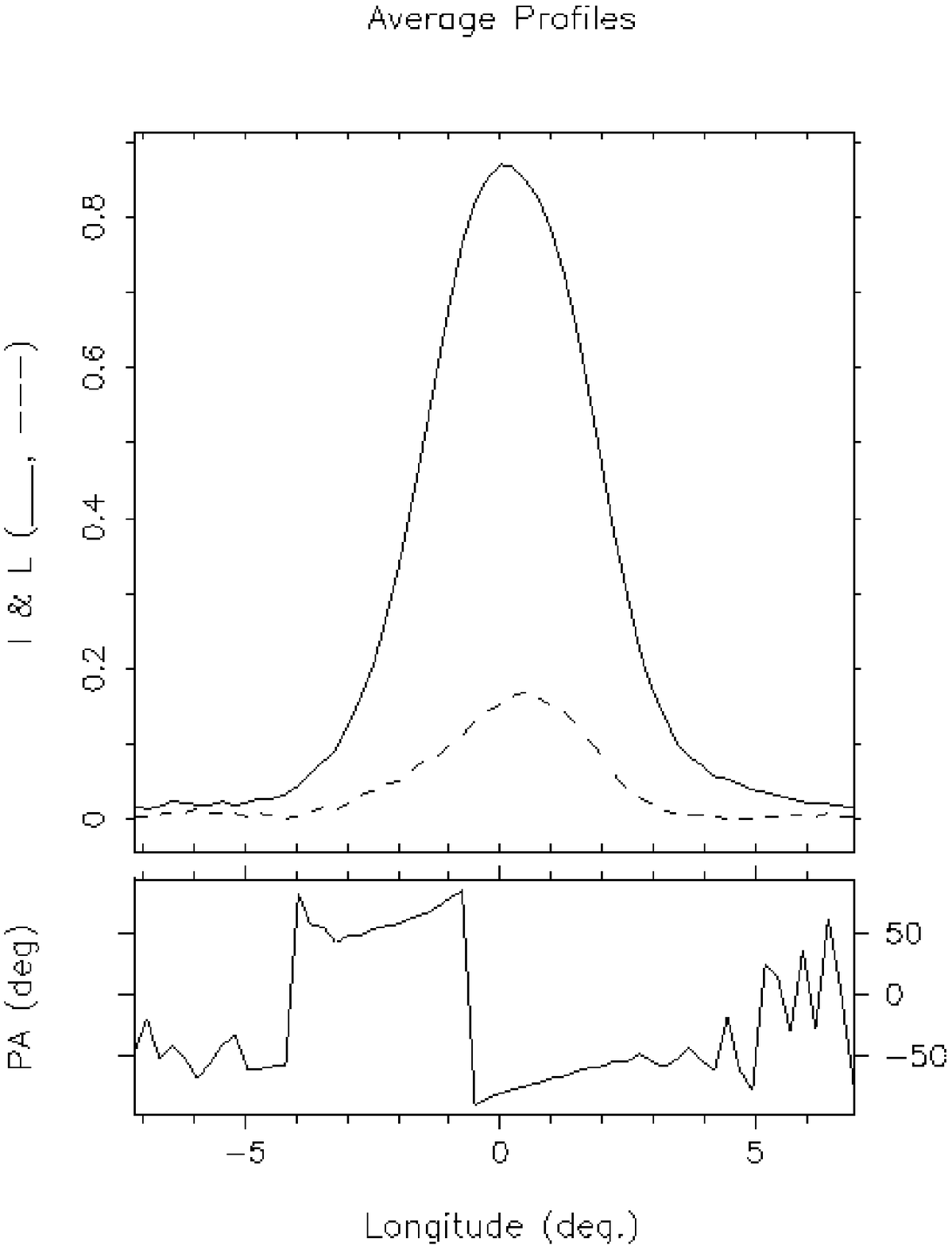}
\caption{Spectral modulation (left) and average linear polarization profile for \mbox{J0837-4135}}
\end{figure}
\begin{figure}
\plottwo{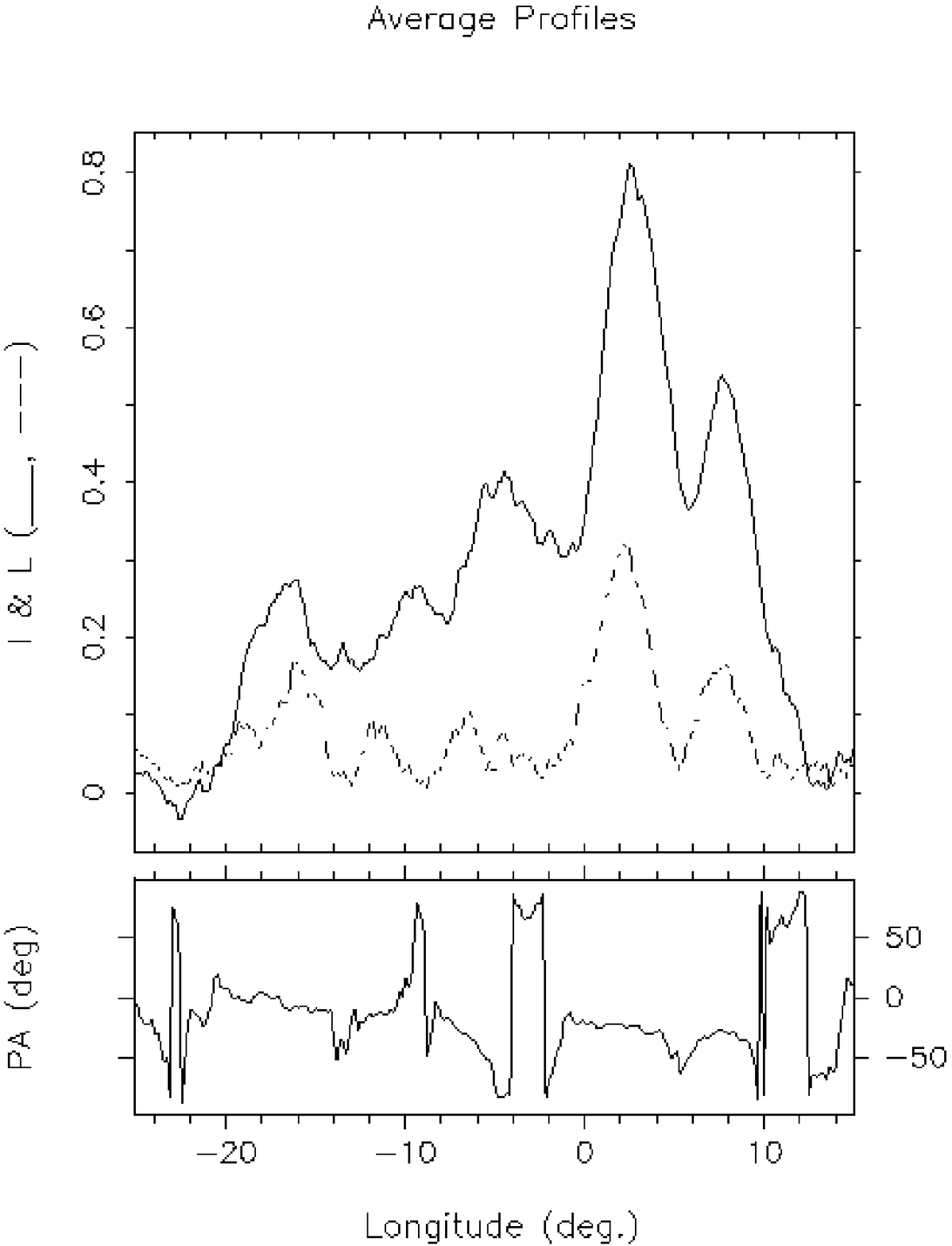}{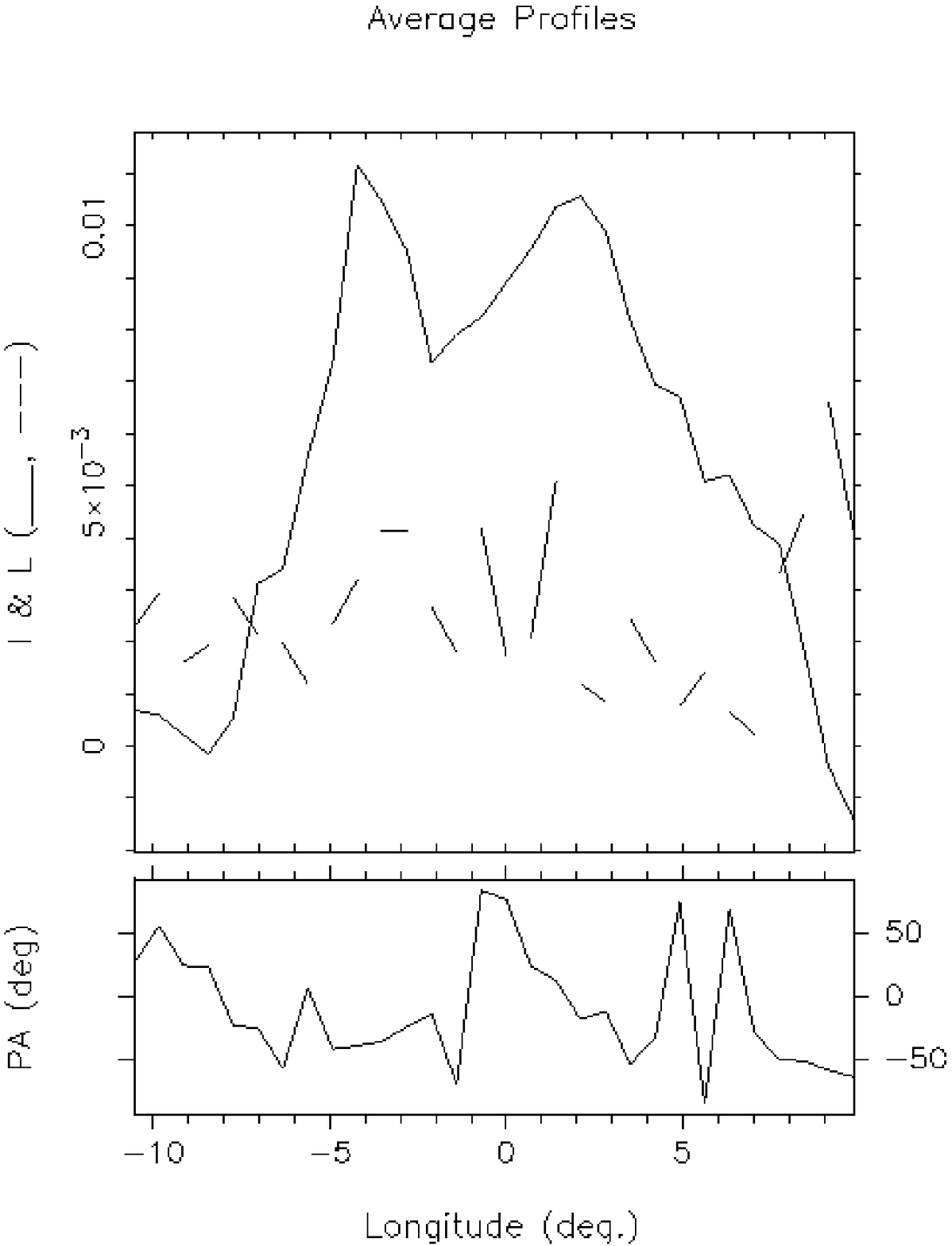}
\caption{The average linear polarization profiles for \mbox{J0846-3533} at 327 MHz (left),
and for B0834+06 at 35 MHz.}
\end{figure}


\begin{references}
\reference Asgekar, A., \& Deshpande, A.A. 1999, elsewhere in this volume
\reference Deshpande, A.A., \& Ramkumar, P. S. 1999, J. Astrophys. Astr., 20, 37 (RD99)
\reference Smirnova, T.V., \& Boriakoff, V. 1997, \aap, 321, 305
\reference Suleimanova, S.A., Volodin, Yu.V., \& Shitov, Yu.P. 1988, \sovast, 32, 177
\end{references}
\end{document}